\documentclass[12pt]{article}
\usepackage{epstopdf}
\usepackage[utf8]{inputenc}
\usepackage[dvips]{color}
\usepackage{epsfig}
\usepackage{amsmath}
\usepackage{hyperref}
\usepackage{graphicx}

\begin{document}
\title{\bf Sound Speed in Extended Chaplygin Fluid}
\author{{Behnam Pourhassan$^{a}$\thanks{Email: b.pourhassan@du.ac.ir}}, {Hoda Farahani$^{a, b}$\thanks{Email: h.farahani@umz.ac.ir}} and Sudhaker Upadhyay$^{c, d, a}$\thanks{Email:  sudhakerupadhyay@gmail.com; sudhaker@associates.iucaa.in}\\
$^{a}${\small {\em  School of Physics, Damghan University, Damghan, Iran}}\\
        {\small {\em P.O.Box 3671641167, Damghan, Iran}}\\
$^{b}${\small {\em  Canadian Quantum Research Center, 204-300232 AveVernon, BCV1T2L7  Canada}}
      \\
$^{c}${\small {\em Department of Physics, K.L.S. College,  Nawada-805110,}}\\
  {\small {\em (a constituent unit of Magadh University, Bodh-Gaya), Bihar, India}}\\
$^{d}${\small {\em  {Visiting Associate, Inter-University Centre for Astronomy
and Astrophysics (IUCAA)}}}\\
  {\small {\em Pune-411007, Maharashtra, India}}   }
  \date{}
\maketitle
\begin{abstract}
\noindent We consider an extended Chaplygin gas equation of state which is driven from D-brane action and construct a cosmological model based on this equation of state.
In this regard, we compute the scale factor of the model under a certain approximation. The conservation equation of this case is a non-linear differential equation which should solve using the special conditions.
We also analyze the stability of the model by using sound speed as well as adiabatic index and discuss certain special cases of the model. We find special equation of state in this model which yields to dynamical and thermodynamical stability. Furthermore, we study the cosmological consequences of this model under certain conditions.
\end{abstract}

\section{Overview and motivation}
Even after confirmation that the most of the Universe filled by dark energy and dark matter, the nature of this dark sector of the Universe  remains
a mystery.  Therefore, determining the dark Universe nature is an important challenge in theoretical physics. In that case, particle physics need to understand elementary particles which constitute the dark energy and dark matter. There are several phenomenological and theoretical models   to describe the accelerating expansion of the Universe (Riess et al., 1998; Perlmutter et al., 1999; Deffayet et al., 2002; Chaubey and Shukla 2013). Most of them are based on emphasis of  the fact that dark energy and cold dark matter have negative pressure. But,
neither  cold dark matter  nor dark energy  has direct observational test to confirm their reality. Therefore, a unified scenario of dark matter and dark energy suggests that these two components
are different aspects of a single fluid as proposed by Matos and Urena-Lopez (2000).\\

One of the interesting models describing dark side of the Universe is based on the Chaplygin gas (CG) fluid (Bento et al., 2002; Kamenshchik et al., 2001). The primary CG model was not consistent with recent observational data like SNIa, BAO, and CMB (Makler, et al., 2003; Sandvik, et al., 2004; Zhu 2004; Bento et al., 2003). Hence, generalized Chaplygin gas (GCG) was proposed, which is indeed a unification of dark energy and dark matter (Bilic et al., 2002; Bazeia 1999). Subsequently, the GCG changed to the modified Chaplygin gas (MCG) (Debnath et al., 2004) to obtain more agreement with recent observations. More extensions  also exist such as generalized cosmic Chaplygin gas (GCCG) which has been done by Gonzalez-Diaz (2003), or modified cosmic Chaplygin gas (MCCG) which has been studied by Pourhassan (2013). Consideration of viscosity in various CG model is also studied with interests (Saadat and B. Pourhassan 2013). Also, some CG models described accelerating expansion of the Universe by taking variable parameters (Salti et al., 2018; Salti et al., 2019). The latest CG model, so-called extended Chaplygin gas (ECG), is proposed to cover barotropic fluid with the quadratic equation of state (Pourhassan and Kahya 2014a; Kahya et al., 2015; Kahya and Pourhassan 2015).\\

Let us systematically summarize the various CG models studied so far. One of the recent cosmological models including a negative pressure is based on the exotic type of a perfect fluid suggests that the Universe filled by the CG to produce accelerating expansion. This model is described by  the following equation of state (EoS) relating  energy density  $\rho$ and pressure $p$  (Pun et al., 2008):
\begin{equation}\label{s1}
p=-\frac{A}{\rho},
\end{equation}
where $A$ is a positive constant. It should be noted that in the natural units, the energy density and pressure are dimensionless quantities. The EoS given by equation (\ref{s1}) was introduced originally by Chaplygin as a suitable model to reflect the lifting force in an airplane (Chaplygin 1904).\\
Further, GCG is described by the following state equation (Bento et al., 2002):
\begin{equation}\label{s3}
p=-\frac{A}{\rho^{\alpha}},
\end{equation}
where $0\leq\alpha\leq1$. This model provides a cosmic evolution from an initial dust-like behavior to late time which is the cosmological constant. The
Chaplygin gas model is relevant in the stabilization of branes in black hole backgrounds (Kamenshchik and Moschella 2000). \\
 The MCG equation of state is given by Debnath et al., (2004)
\begin{equation}\label{s4}
p=A_{1}\rho-\frac{A}{\rho^{\alpha}},
\end{equation}
 where $A$ and $\alpha$ are positive constants, while $A_{1}$ may be positive or negative constant.
 This equation of state discusses radiation era  at one extreme for  negligibly small
scale factor while a $\Lambda$CDM  model at the other extreme.
 In fact,   $A$ or $A_{1}$ is also considered as a variable (Guo and Zhang 2007).
 A recent equation of state so-called ECG are also obtained  as (Pourhassan and Kahya 2014b)
\begin{equation}\label{s5}
p=\sum_{i=1}^{n}A_{i}\rho^{i}-\frac{A}{\rho^{\alpha}}.
\end{equation}
There is a possibility to write a much more comprehensive equation of state describing viscous MCCG  as follows
\begin{equation}\label{s2}
p=\sum_{i=1}^{n}A_{i}\rho^{i}-\frac{1}{\rho^{\alpha}}\left[\frac{A}{1+w} +\left(\rho^{1+\alpha}-\frac{A}{1+w}+1\right)^{-w}-1\right]-\Pi,
\end{equation}
where $w$ is called cosmic parameter \cite{13} and $\Pi$ corresponds to viscosity which is generally depends on the energy density and can be written as powers of $\rho$ ($\Pi\propto \rho^{m}$). All the equations of state (\ref{s1})-\ref{s5}) are particular cases of (\ref{s2}). For example, it is easy to check that for $\Pi=w=A_{n}=0$ and $\alpha=1$ (\ref{s2}) reduces to the equation (\ref{s1}).
For $\Pi=w=A_{n}=0$, the equation (\ref{s2}) coincides with (\ref{s3}).
For $\Pi=w=0$ and $n=1$   (\ref{s2}) reduces to (\ref{s4}).
Eq. (\ref{s5}) can be obtained from   (\ref{s2}) by setting $w=0$.\\

A more comprehensive equation of state is  obtained from string theory by Pourhassan (2019)
\begin{equation}\label{s6}
p=\sum_{i\in R}A_{i}\rho^{i},
\end{equation}
where $i$ may have a positive, negative, integer, and non-integer number. The same procedure already has been considered by Ogawa (2000) for the equation of state (\ref{s1}).

Our analysis is based on this particular   equation of state (\ref{s6}).
We unify all of the mentioned CG equations of state using the Nambo-Goto action for a $d$-brane moving in a ($d+2$)-dimensional space-time. In this regard, we first consider several CG equations of state and solve the string equation of motion in order to obtain a general equation of state which generates all of the above mentioned equations. The next purpose of the letter is to study the cosmological consequence of this model. Therefore, we try to obtain the relation between energy density and scale factor.  In this regard,    following the conservation law, we estimate the scale factor of the model. We  consider  first-order approximation here and, as per expectation, we find that the scale factor reduces with increasing energy
density. This justifies the consistency of our cosmological model.
In order to explain the accelerating expansion of the Universe and describe dark matter effects,
  the equation of state parameter is also calculated.
Furthermore, we derive sound speed  for this modified cosmological model.   We
show that the barotropic case of this model is stable. We check the stability of the model make sure that squared sound speed must be positive. We confirm that the viscous modified Chaplygin fluid model  is a stable model while there are various  unstable  generalized Chaplygin fluid models. By studying sound speed analysis, the model imposes constrain on parameter.

The paper is presented as follows. In Sec. \ref{sec1}, we discuss the scale factor of the model. The sound speed is studied in section \ref{sec2}. A particular case is realized in section \ref{sec3}. Finally, we summarize the results with concluding remarks in the last section.

\section{Scale factor}\label{sec1}
In order to find cosmological implication, we first write conservation law for the fluid with an energy density $\rho$ and pressure $p$ as follows
the conservation law
\begin{equation}\label{s16}
\dot{\rho}+3H(p+\rho)=0,
\end{equation}
where Hubble expansion parameter $H$  is defined in terms of   the scale factor  $a(t)$ by
\begin{equation}\label{s17}
H=\frac{\dot{a}}{a}.
\end{equation}
The conservation equation (\ref{s16})  for the pressure  (\ref{s6})  and Hubble parameter
(\ref{s17}) takes the following form:
\begin{equation}\label{s18}
d\rho+3\frac{d{a}}{a}\left(\rho+\sum{A_{i}\rho^{i}}\right)=0.
\end{equation}
In order to obtain an analytical solution, we assume the same value for all coefficients and the following expansion,
\begin{equation}\label{s19}
\sum_{i=-m}^{n}{A_i\rho^{i}}=\frac{A}{\rho-1}(\rho^{n+1}-\rho^{-m}),
\end{equation}
where we assumed the same value for all coefficients. In that case, the expression
(\ref{s18}) reduces to
\begin{equation}\label{s20}
 \ln{\frac{a}{a(0)}}=-\frac{1}{3}\int{\frac{d\rho}{\rho+\frac{A}{\rho-1}(\rho^{n+1}-\rho^{-m})}}.
\end{equation}
The solution for above equation for the special case of $m=1$ and $n=3$ is computed by
\begin{equation}\label{s21}
 \ln{\frac{a}{a(0)}} =C_{-}\tan^{-1}\left(\frac{{\mathcal{B}}_{+}}{{\mathcal{A}}_{-}}\right) -C_{+}\tan^{-1}\left(\frac{{\mathcal{B}}_{-}}{{\mathcal{A}}_{+}}\right),
\end{equation}
where
\begin{eqnarray}\label{s22}
{\mathcal{A}}_{\pm}&=&\sqrt{10A^{2}\pm2A\sqrt{A(5A-4)}+4A},\nonumber\\
{\mathcal{B}}_{\pm}&=&A(1+4\rho)\pm\sqrt{A(5A-4)},\nonumber\\
C_{+}&=&\frac{4}{3{\mathcal{A}}_{+}}\sqrt{\frac{A}{5A-4}},\nonumber\\
C_{-}&=&\frac{4}{3{\mathcal{A}}_{-}}\sqrt{\frac{A}{5A-4}}.
\end{eqnarray}
Hence, the scale factor is given by
\begin{equation}\label{s23}
a=a(0)\exp\left[C_{-}\tan^{-1}\left(\frac{{\mathcal{B}}_{+}}{{\mathcal{A}}_{-}}\right)-C_{+}\tan^{-1}\left(\frac{{\mathcal{B}}_{-}}{{\mathcal{A}}_{+}}\right)\right].
\end{equation}
At the first order approximation (for small $\rho$ we neglect $\mathcal{O}(\rho^{2})$), we obtain the late time scale factor as
\begin{equation}\label{s25}
a=a(0)\exp\left[C_{0}-C_1\rho\right],
\end{equation}
where
\begin{eqnarray}\label{s27}
C_{0}&=&C_{-} \tan^{-1}\left(\frac{A+\sqrt{A(5A-4)}}{\sqrt{10A^{2}-2A\sqrt{A(5A-4)}+4A}}\right) \nonumber\\
 &-&C_{+} \tan^{-1}\left(\frac{A-\sqrt{A(5A-4)}}{\sqrt{10A^{2}+2A\sqrt{A(5A-4)}+4A}}\right),\nonumber\\
C_1 &=&C_{+}\frac{4A}{\sqrt{10A^{2}+2A\sqrt{A(5A-4)}+4A}\left(1+\frac{(A-\sqrt{A(5A-4)})^{2}}{\sqrt{10A^{2}+2A\sqrt{A(5A-4)}+4A}}\right)}\nonumber\\
&-&C_{-}\frac{4A}{\sqrt{10A^{2}-2A\sqrt{A(5A-4)}+4A}\left(1+\frac{(A+\sqrt{A(5A-4)})^{2}}{\sqrt{10A^{2}-2A\sqrt{A(5A-4)}+4A}}\right)}.
\end{eqnarray}

\begin{figure}[h!]
 \begin{center}$
 \begin{array}{cccc}
\includegraphics[width=80 mm]{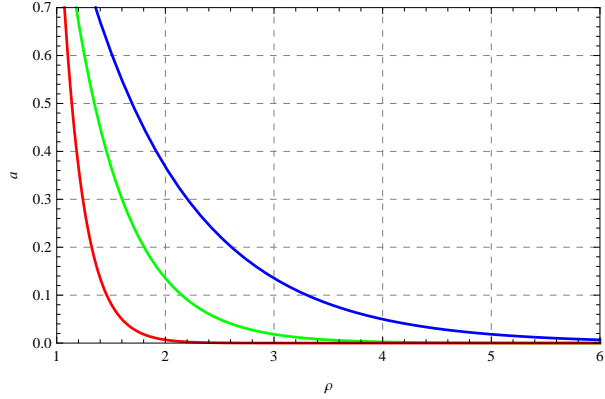}
 \end{array}$
 \end{center}
\caption{Typical behavior of scale factor with $\rho$ for $a({0})=1$. Here, $ C_{0}=C_1 = 1$ case is denoted by blue line, $ C_{0}=C_1 = 2$ case is denoted by
green line and $ C_{0}=C_1= 5$ case is denoted by red line.}
 \label{1}
\end{figure}

In order to study the dependence of scale factor on energy density, we plot a graph of the scale factor given by the equation (\ref{s23}) in Fig.
\ref{1}. From the plot, it is obvious that the  scale factor  reduces with increasing energy density  as expected. This establishes the consistency of our  cosmological model based on the equation of state (\ref{s6}). Therefore, we can say  that the Universe is filled by a fluid described by  the following equation of state:
\begin{equation}\label{s28}
p=\frac{A}{\rho-1}(\rho^{n+1}-\rho^{-m}).
\end{equation}
Therefore, we have the following equation of state parameter:
\begin{equation}\label{s29}
\omega=\frac{p}{\rho}=\frac{A}{\rho-1}(\rho^{n}-\rho^{-m-1}).
\end{equation}
This explains the accelerating expansion of the Universe and describes dark matter effects as well. It can be shown by computing the deceleration parameter
\begin{equation}\label{q}
q=-\frac{a\ddot{a}}{{\dot{a}}^{2}}.
\end{equation}
Using the solution (\ref{s23}) we can obtain a negative deceleration parameter at a late time. However, there is a situation where the deceleration/acceleration phase transition happens.
Next, we study sound speed in such fluid to analyze the stability of the model.

\section{Sound speed}\label{sec2}
\begin{figure}[h!]
 \begin{center}$
 \begin{array}{cccc}
\includegraphics[width=60 mm]{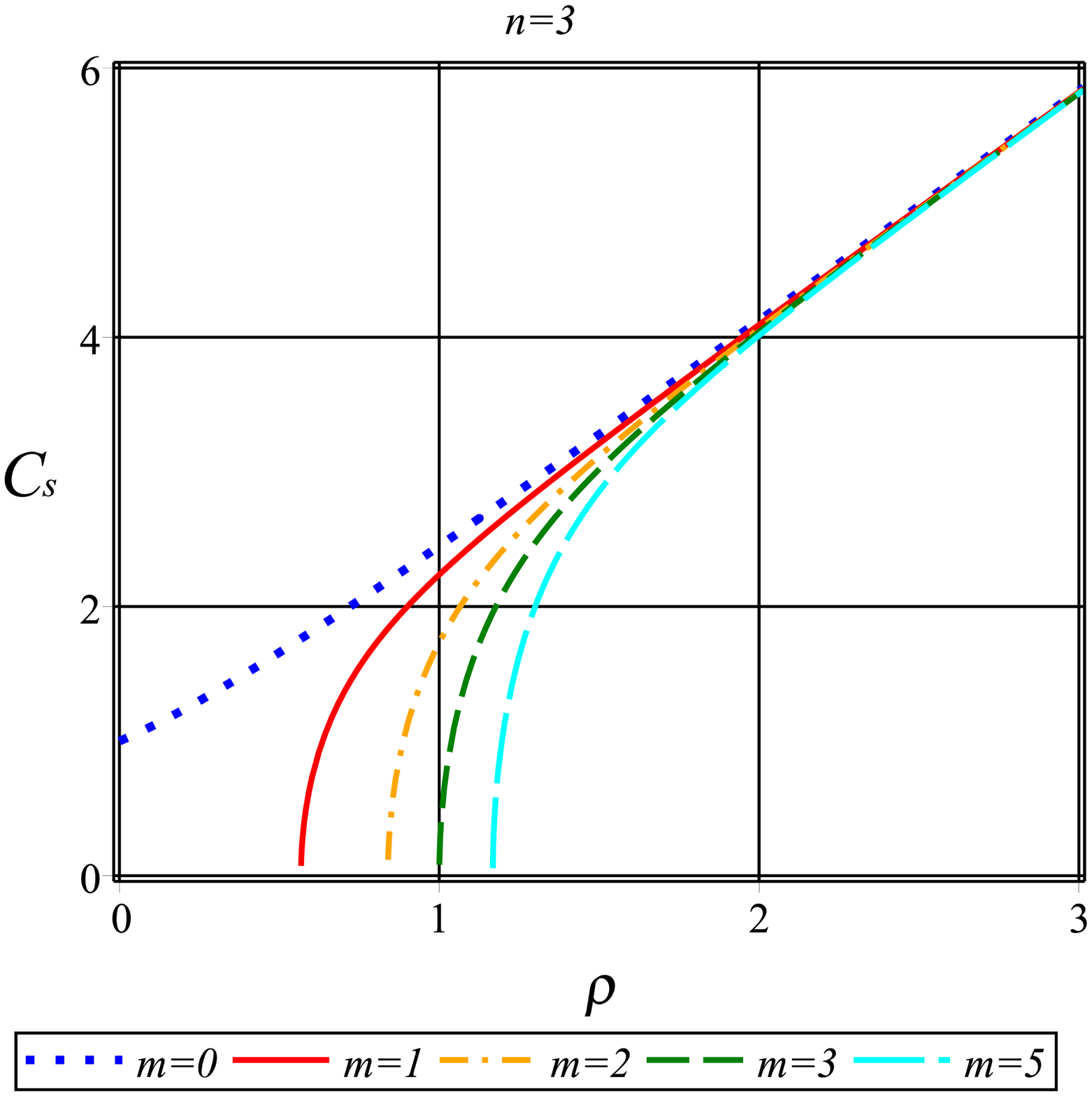}\includegraphics[width=60 mm]{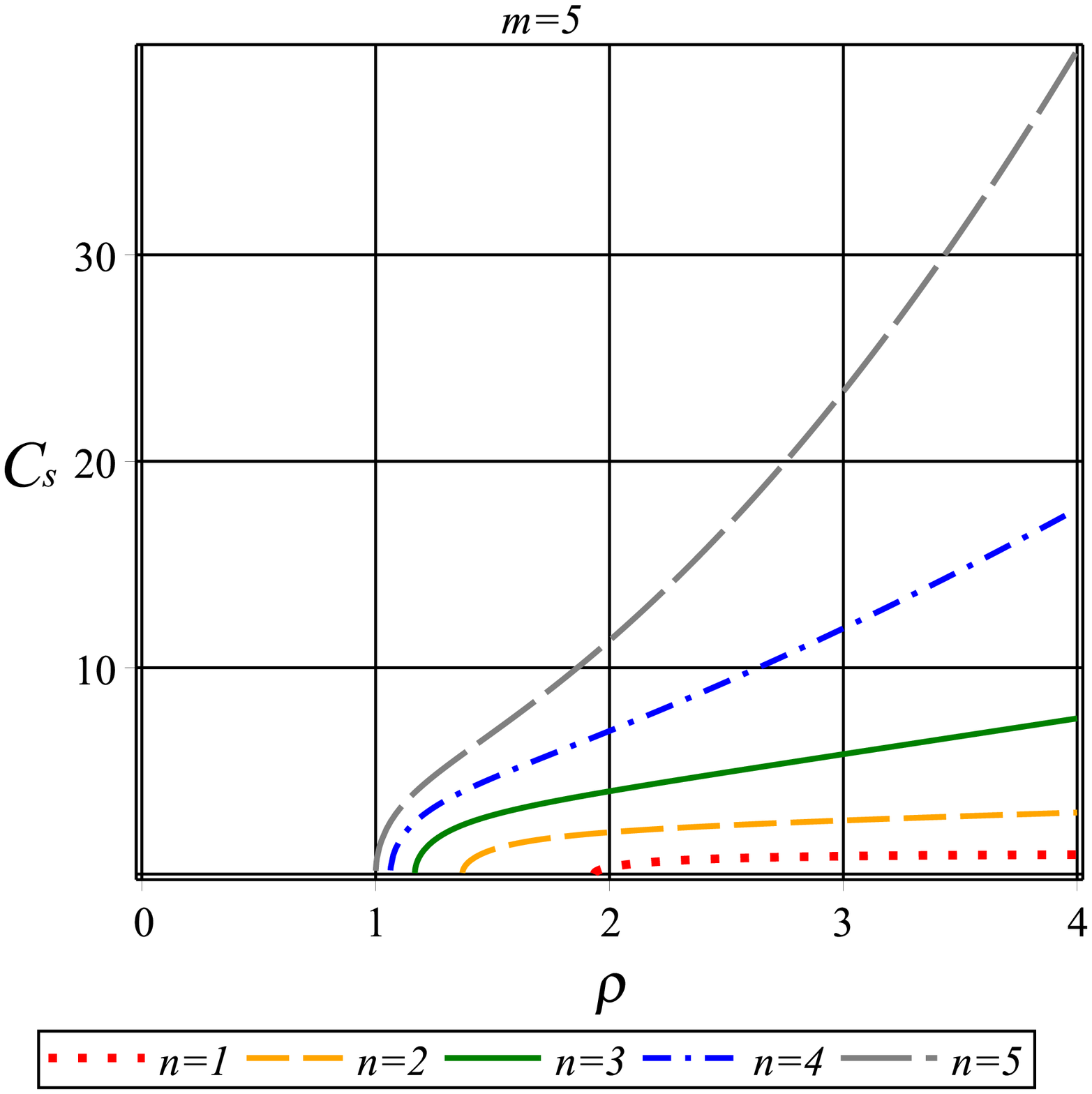}\\
\includegraphics[width=60 mm]{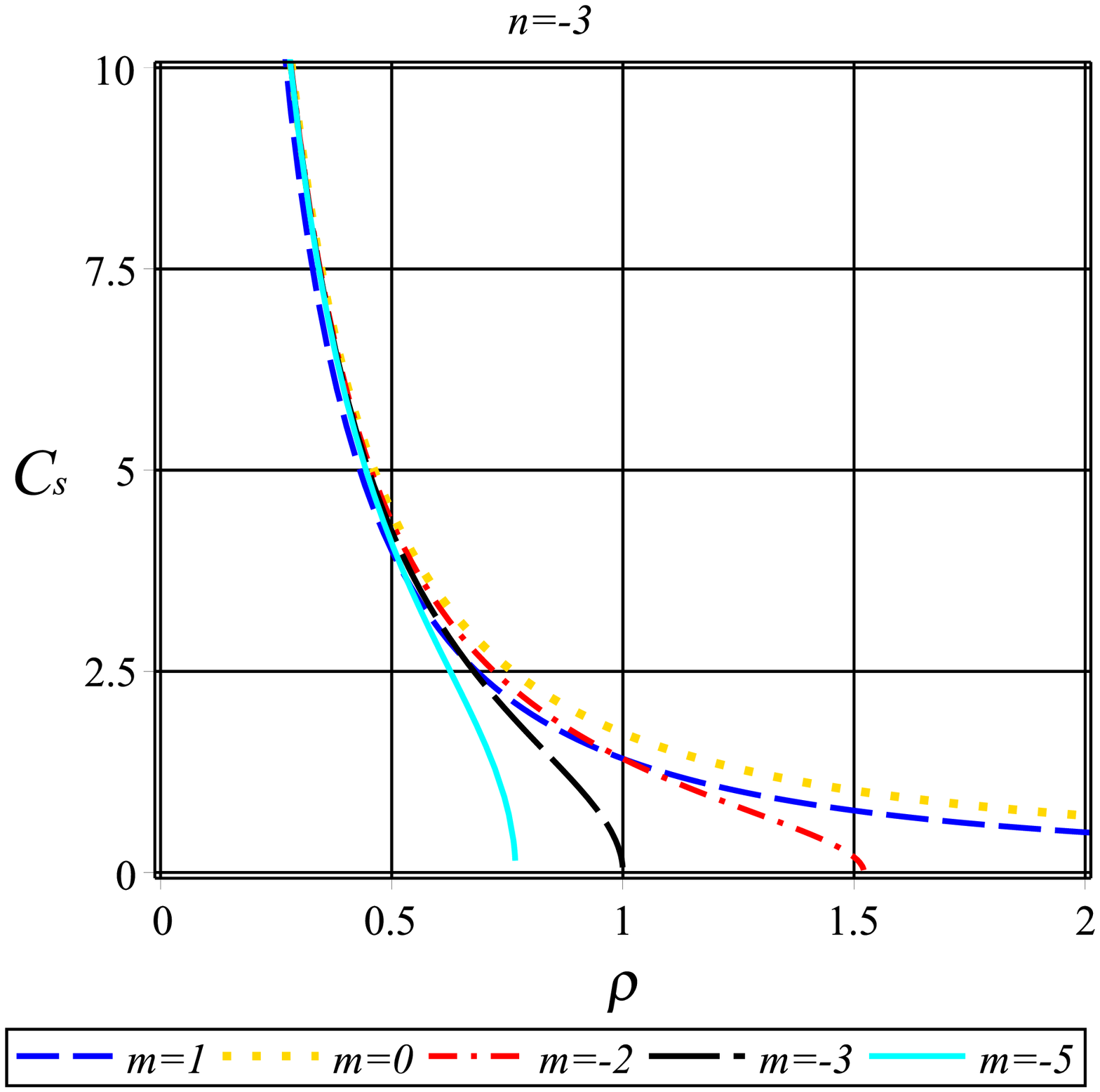}\includegraphics[width=60 mm]{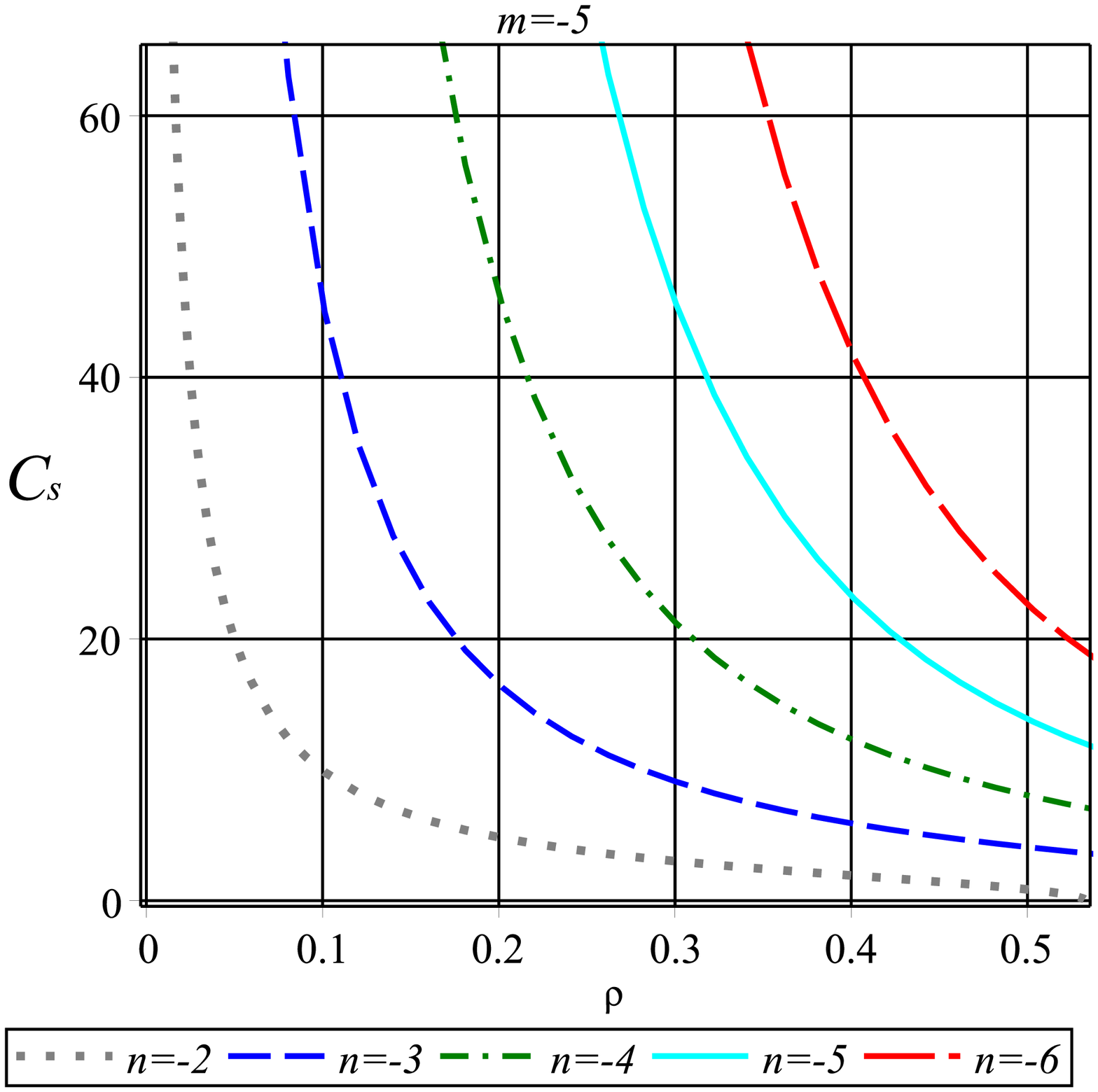}
 \end{array}$
 \end{center}
\caption{Sound speed in terms of energy density for $A=1$.}
 \label{C}
\end{figure}

In this section, we analyze the sound speed of the fluid to discuss the stability of the model.
The sound speed for the cosmic fluid model, $C_{s}$, can be estimated from the following relation:
\begin{equation}\label{s30}
C_{s}^{2}=\frac{dp}{d\rho}.
\end{equation}
Corresponding to equation of state (\ref{s28}), this formula yields
\begin{equation}\label{s31}
C_{s}^{2}=A\frac{(n\rho-n-1)\rho^{n}+((m+1)\rho-m)\rho^{-m-1}}{(\rho-1)^{2}}.
\end{equation}
The sound speed should be positive in the stable model and hence this is justified by
first consideration that constant $A$ must be positive as mentioned early in Eq. (\ref{s1}). In the plots of Fig. \ref{C} we can see behavior of sound speed. Red solid line shows special case of $m=1$ and $n=3$ which discussed in previous section. For several values of positive $m$ and $n$ we see real sound speed which is increasing function of energy density. However, there are situations with negative $m$ or $n$ where sound speed is decreasing function of energy density. It help us to find dynamically stable model.

\begin{figure}[h!]
 \begin{center}$
 \begin{array}{cccc}
\includegraphics[width=55 mm]{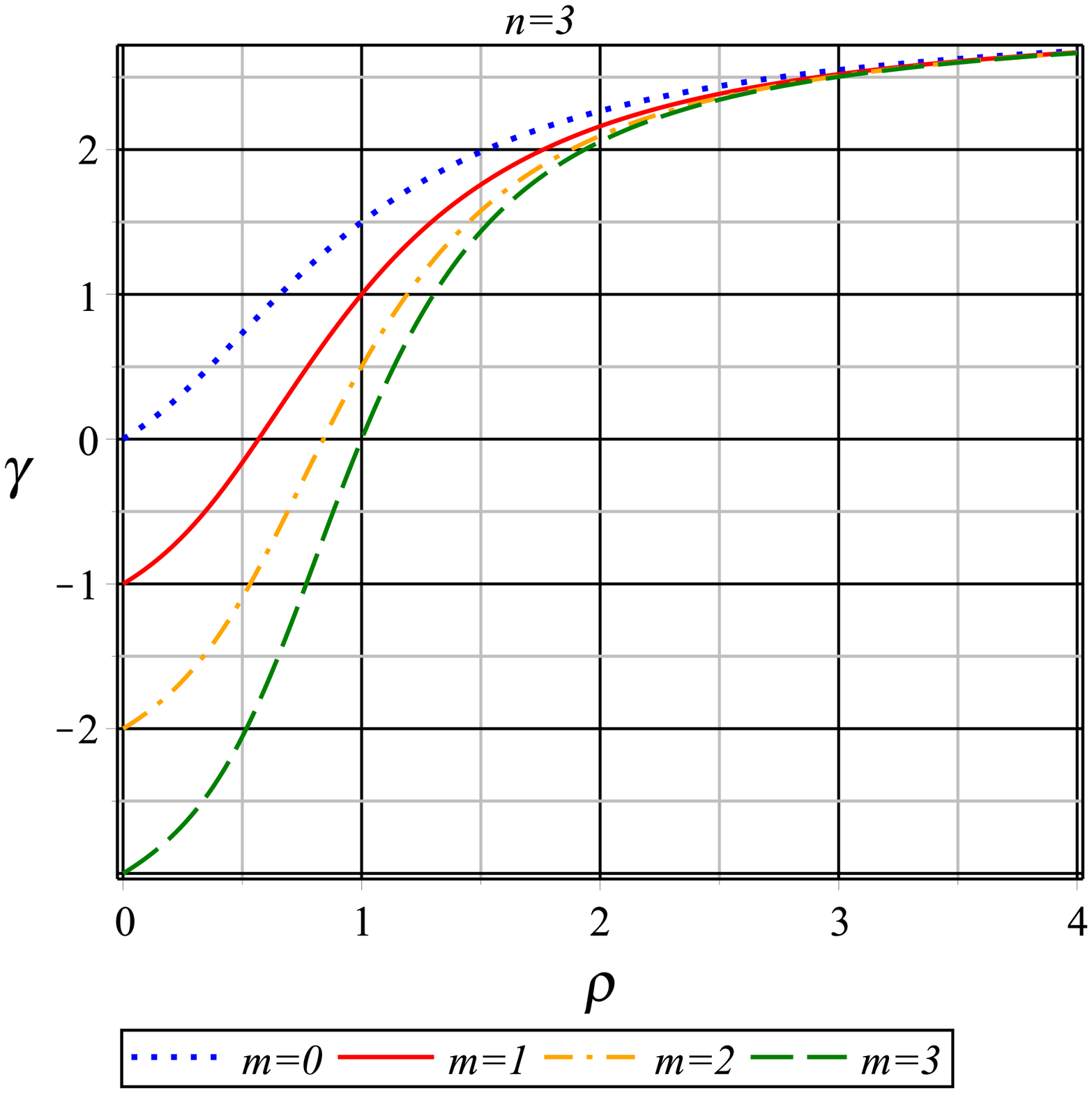}\includegraphics[width=55 mm]{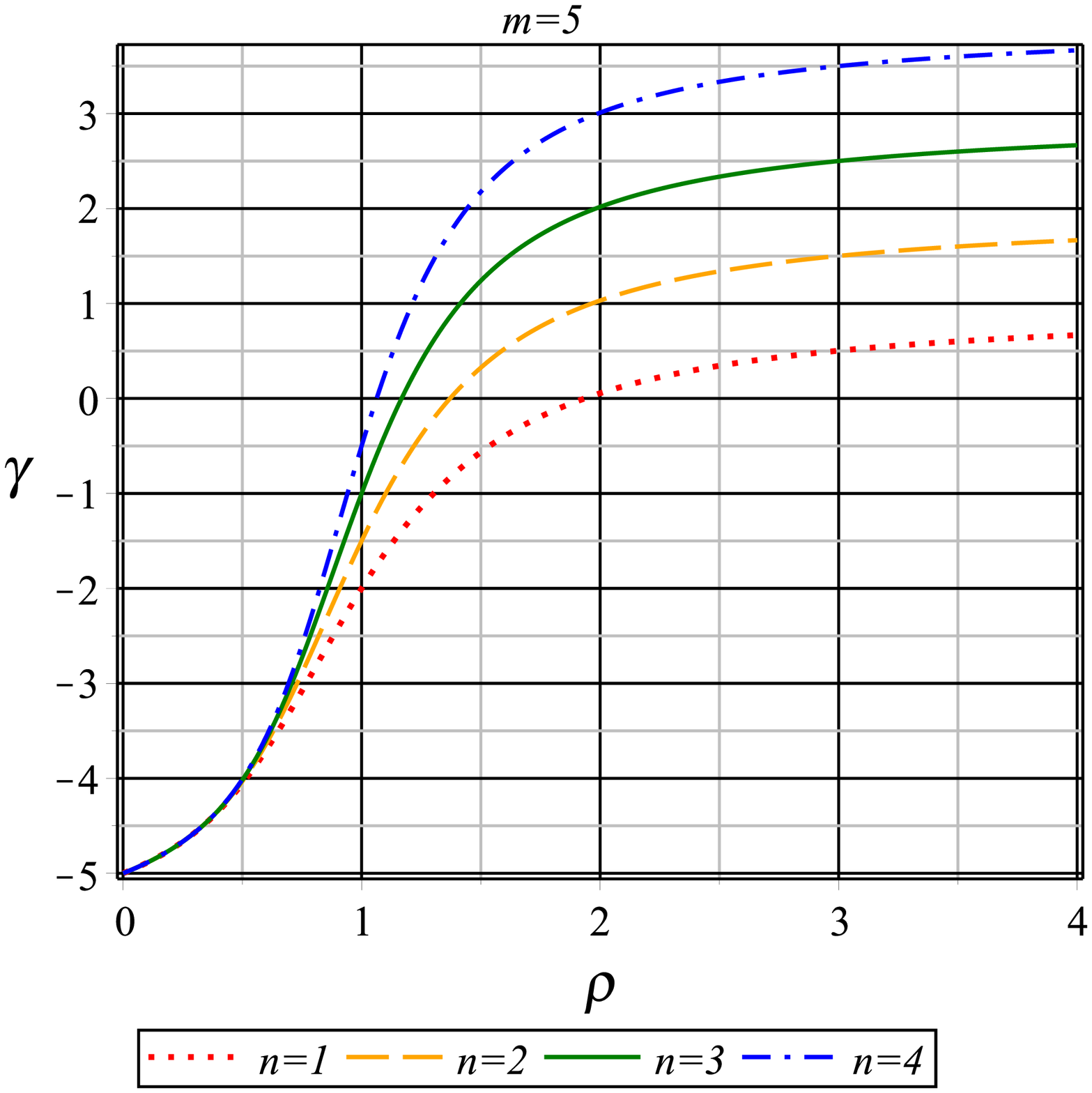}\includegraphics[width=55 mm]{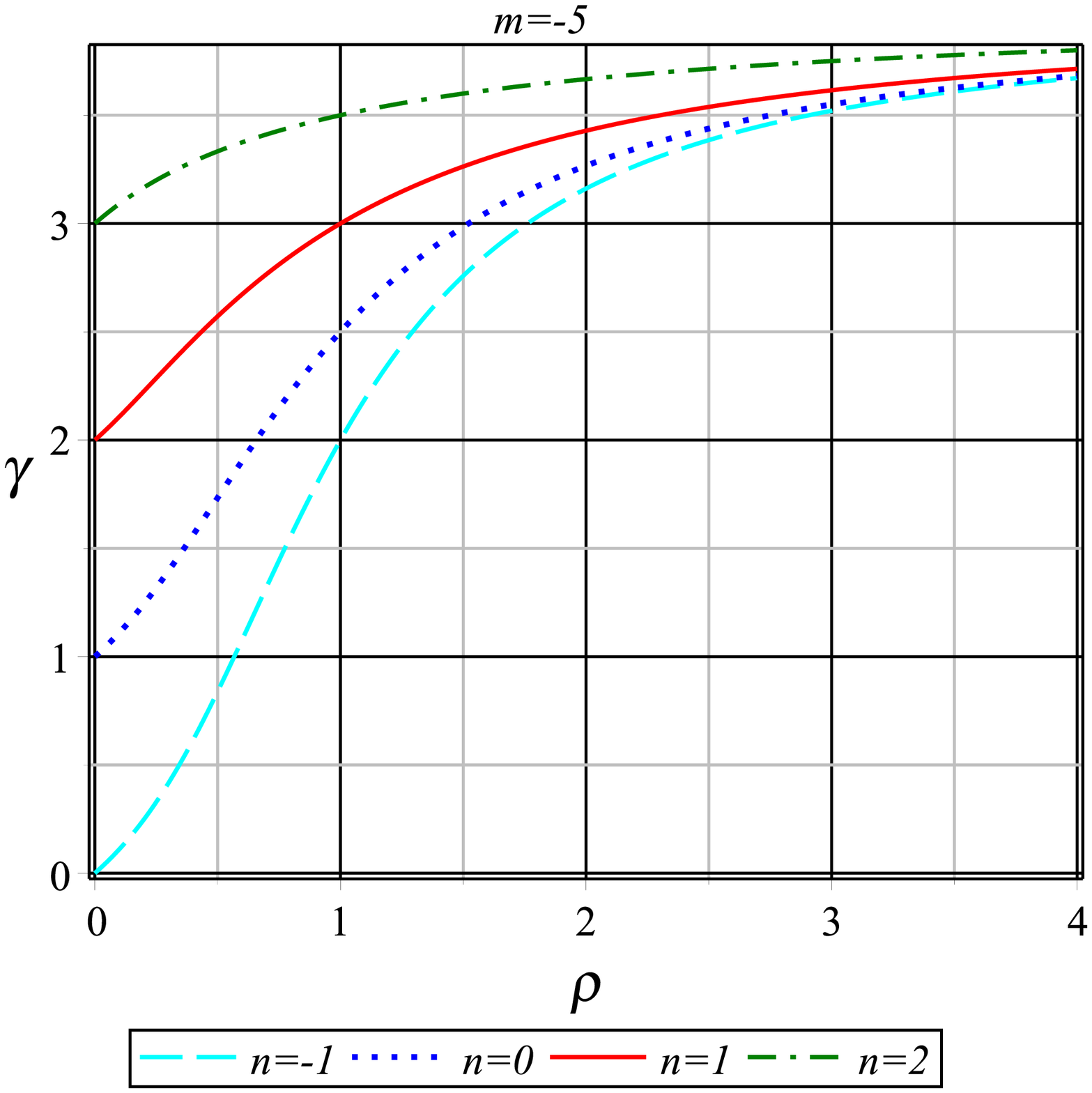}
 \end{array}$
 \end{center}
\caption{Adiabatic index in terms of energy density for $A=1$.}
 \label{a}
\end{figure}

In order for establish a physical stable and acceptable model we need to study also thermodynamical stability. It may be found by analyzing the adiabatic index,
\begin{equation}\label{a1}
\gamma=\frac{C_{p}}{C_{v}},
\end{equation}
where $C_{p}$ and $C_{v}$ are specific heat at constant pressure and constant volume respectively. For the general fluid, it yields to the following relation at the constant entropy,
\begin{equation}\label{a2}
\gamma=\left(\frac{\partial \ln{p}}{\partial \ln{\rho}}\right)_{S}.
\end{equation}
It is estimated that the value of this parameter must be greater than $frac{4}{3}$ for a dynamically stable model. In that case, using the equation of state (\ref{s28}) one can obtain,
\begin{equation}\label{a2}
\gamma=\frac{(n(1-\rho)+1)\rho^{n+1}-((m+1)\rho-m)\rho^{-m}}{(\rho-1)(\rho^{-m}-\rho^{n+1})}.
\end{equation}
In plots of Fig. \ref{a} we draw adiabatic index in terms of energy density. For some cases of positive $m$ and $n$ we find that the condition $\gamma\geq\frac{4}{3}$ satisfied at the early time with the large energy density. For example, in the case of $n=3$ we find $\gamma\approx\frac{8}{3}$ at the early time which decreased by time. On the other hand we can see the complete stable model for some positive $n$ and negative $m$ (see the last plot of Fig. \ref{a}).

Now, we consider the case of early time where $\rho\gg1$ and in this case the first term of the numerator of (\ref{s31}) is dominant and hence the model is dynamically stable for $n\geq0$. On the other hand, at the late time where $\rho\ll1$ the second term of the numerator in the equation (\ref{s31}) is dominant and the model is only stable for $m\leq0$. Hence, the scale factor (\ref{s25}) which is obtained for $n=3$ and $m=1$ leads  to an instability  to the model for the late time. This motivates us we consider a very special case to construct a suitable cosmological model in the next section. At the moment, we assume $m=-n$ in expression (\ref{s31}) which results
\begin{equation}\label{s32}
C_{s}^{2}=\frac{nA(\rho^{n+2}+\rho^{n}-2\rho^{n+1})}{\rho(\rho-1)^{2}}.
\end{equation}
From the above expression it is obvious that the sound speed vanishes in the case of $n=0$. Hence, the requirement for non-vanishing sound speed here is that $n>0$. For positive integer valued $n$, we find that $C_{s}^{2}\geq A$, here equality holds for $n=1$.\\
In that case ($m=-n$ with $n>0$), we find that
\begin{equation}\label{a3}
\gamma\approx n,
\end{equation}
hence, for the all cases of $n\geq\frac{4}{3}$ we have completely stable model.
Therefore, the equation of state in the stable model may read as,
\begin{equation}\label{s33}
p=\frac{A}{\rho-1}(\rho^{n+1}-\rho^{n}),
\end{equation}
with $n=1, 2, 3, \cdots$, which may reduced to barotropic equation of state ($p=A\rho^{n}$) which already studied deeply in literature.
\section{Very special case}\label{sec3}
Now, we consider only three non-zero coefficient $A_{1}=A$, $A_{-\alpha}=-B$ and $A_{\frac{1}{3}}=-\frac{\xi}{\sqrt{3}}$, where $B$ is a positive constant and $\xi$ is a constant viscous coefficient (Khadekar et al., 2019). Hence, we recover viscous modified Chaplygin gas with the following equation of state:
\begin{equation}\label{s34}
p=A\rho-\frac{B}{\rho^{\alpha}}-\frac{\xi}{\sqrt{3}}\rho^{\frac{1}{3}},
\end{equation}
where $\Pi=\xi H=\frac{\xi}{\sqrt{3}}\rho^{\frac{1}{3}}$ is used.
In the case of $\alpha=\frac{1}{3}$ and assuming $X\equiv\frac{1}{\rho^{\frac{2}{3}}}$ one can obtain the sound speed
\begin{equation}\label{s35}
C_{s}^{2}=\frac{B}{3}X^{2}-\frac{\xi\sqrt{3}}{9}X+A.
\end{equation}
Here, we observe that  $C_{s}^{2}>0$ if
\begin{eqnarray}
 A \geq \frac{\xi^2}{36B}.
\end{eqnarray}
The above relation is a required condition for  stability of the model.

\begin{figure}[h!]
 \begin{center}$
 \begin{array}{cccc}
\includegraphics[width=83 mm]{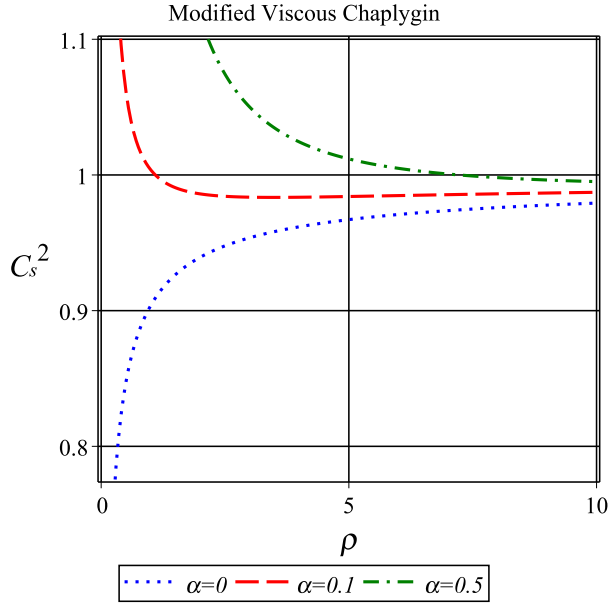}
 \end{array}$
 \end{center}
\caption{Square sound speed in terms of energy density for $A=B=1$ and $\xi=0.5$ by variation of $\alpha$.}
 \label{2}
\end{figure}

In order to study the stability for the general case, we draw square sound speed (\ref{s35})    with respect to $\rho$ in Fig. \ref{2}. Here, we observe   that the model is completely stable for all parameter values in the range $0\leq A\leq2$, $0\leq\xi\leq1$, $0\leq\alpha\leq1$ and $0\leq B\leq2$.

\begin{figure}[h!]
 \begin{center}$
 \begin{array}{cccc}
\includegraphics[width=83 mm]{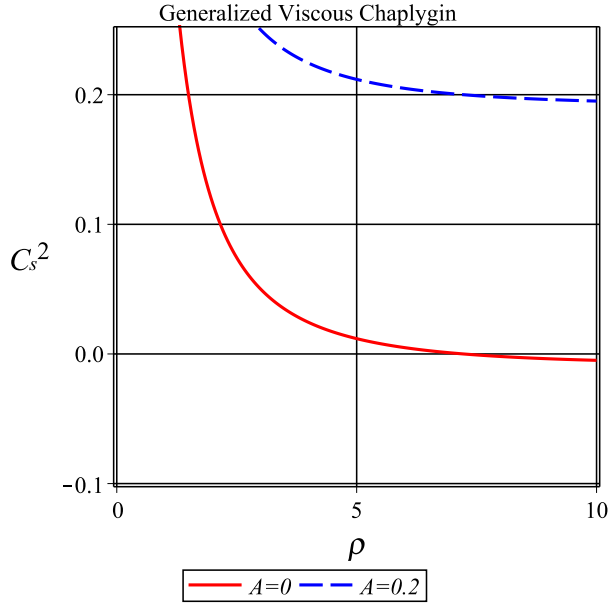}
 \end{array}$
 \end{center}
\caption{Square sound speed in terms of energy density for $\alpha=0.5$, $B=1$ and $\xi=0.5$ by variation of $A$.}
 \label{3}
\end{figure}

In the case of $\alpha=0$ where we have viscous barotropic fluid, the sound speed is increasing function of energy density and, hence, it is decreasing function of time. For other cases with $\alpha>0$, the sound speed is increasing function of time (decreasing by energy density) which diverges at a late time. Here, one can conclude that in all cases the sound speed is a constant at the early time. As per expectation, the viscous coefficient decreases the value of sound speed. Another important result can be seen for the case of $A=0$ which yields to negative $C_{s}^{2}$ at an early time (which is illustrated by Fig. \ref{3}). It means that various versions of generalized Chaplygin gas may be unstable at the late time.
In the Fig. \ref{3}, we draw squared sound speed of the GCG models to show that model yields to imaginary sound speed at the early time.
\section{Discussions and conclusions}\label{sec4}
We have considered a model for CG which is inspired from the string theory. This model is used to unify   dark energy and dark matter to describe accelerating expansion of the Universe which is in agreement with the recent observational data. First, following from the conservation law, we have obtained the scale factor of the model. Here, we have considered first-order approximation.  As per expectation, we have found that the scale factor reduces with increasing energy
density and this establishes the consistency of our cosmological model.
In order to explain  accelerating expansion of the Universe and describe dark matter effects,
 we have calculated the equation of state parameter as well.

Furthermore, we have studied sound speed in this modified extended Chaplygin gas model.   We
have found that the barotropic-like case of this model is stable. We studied dynamical stability of this model by analyzing sound speed via the fact that squared sound speed must be positive. Also we studied thermodynamical stability of the model by analyzing the adiabatic index and found that for the all cases of $m=-n$ with $n\geq\frac{4}{3}$ we have completely stable model. We found that the viscous modified Chaplygin gas is the stable model while various versions of generalized Chaplygin gas is unstable model. We constrained the model parameter by doing sound speed analysis and found that special limit on $A$ parameter is such that $C_{s}^{2}>A$.
Certainly,  we are lacking a general solution for the equation (\ref{s20})   which is the goal of our future investigations in order to construct more comprehensive cosmological model.\\\\

\end{document}